\documentclass[aps,prl,twocolumn,superscriptaddress,showpacs]{revtex4}
\usepackage{amsmath}
\usepackage{graphicx}
\usepackage{epsfig}
\usepackage{bm}

\def\be{\begin{equation}}
\def\ee{\end{equation}}
\def\bea{\begin{eqnarray}}
\def\eea{\end{eqnarray}}
\def\bean{\begin{mathletters}\begin{eqnarray}}
\def\eean{\end{eqnarray}\end{mathletters}}

\begin{document}

\title{Exact Solution for 1D Spin-Polarized Fermions with Resonant Interactions}

\author{Adilet Imambekov}
\affiliation{Department of Physics and Astronomy, Rice
University, Houston, Texas 77005, USA}
\author{Alexander A. Lukyanov}
\affiliation{Abingdon Technology Centre, Schlumberger, Abingdon,
OX14 1UJ, UK}
\author{Leonid I. Glazman}
\affiliation{Department of Physics, Yale University, New Haven,
Connecticut 06520, USA}
\author{Vladimir Gritsev}
\affiliation{Physics Department, University of Fribourg, Chemin
du Musee 3, 1700 Fribourg, Switzerland}

\begin{abstract}
Using  the asymptotic Bethe Ansatz, we obtain an exact solution
of the many-body problem for 1D spin-polarized fermions
with resonant p-wave interactions, taking into account the
effects of both scattering volume and effective range. Under
typical experimental conditions, accounting for the effective
range, the properties of the system are significantly modified 
due to the existence of ``shape'' resonances.
The excitation spectrum of the considered model has 
unexpected features, such as the inverted position of
the particle- 
and hole-like branches at small momenta, and
roton-like minima. We  find that the frequency of the
``breathing'' mode  in the harmonic trap 
provides an unambiguous signature of the effective range.

\end{abstract}

\date{\today}
\maketitle

Experimental progress in the cooling and trapping of ultracold
atomic gases makes it possible to investigate their properties
under strong transverse confinement, when the motion of atoms is
effectively one-dimensional (1D). In such experiments the 1D
interaction parameters are precisely known and can be tuned using
Feshbach resonances~\cite{1} or by varying the harmonic
transverse confinement  strength. Recent
experiments~\cite{KWW,paredes} have allowed for parameter-free
comparison of 1D Bose gas properties with a theoretical
description based on the exactly solvable Lieb-Liniger (LL)
model~\cite{LL}. The possibility to compare experimental results
with the outcomes of many-body calculations  revived an interest
in the field of exactly solvable 1D systems:  spin-1/2
fermions~\cite{FRZ}, Bose-Fermi
mixtures~\cite{mixture1,mixture2}, and spinor
bosons~\cite{spinor_b} and fermions~\cite{spinor_f}.


 In this Letter, we obtain an exact solution  for 1D
spin-polarized fermions under resonant scattering
conditions~\cite{pricoupenko1}, which is 
relevant for $^{40}$K and $^6$Li atoms near p-wave Feshbach
resonances~\cite{p-wave1,p-wave2}. Such a 1D experimental system
has been realized for $^{40}$K~\cite{Gunter}. The related 3D
problem has also received significant theoretical attention
recently~\cite{3D}. For spin-polarized  fermions only scattering
in odd partial wave channels is present, and at low energies,
p-wave scattering is the strongest.  If only the "scattering
volume" is taken into account and  the "effective range" of
p-wave scattering is neglected (see Eq. (1) for definitions),
then the projection to 1D~\cite{GB} results in  a fermionic
Cheon-Shigehara  (CS) model~\cite{CSh}, which is dual to the
bosonic LL model. For such a  model, strongly interacting
fermions with resonant interactions are mapped to weakly
interacting  bosons, the so-called fermionic Tonks-Girardeau
(fTG) limit~\cite{GW}. However, it was shown 
by L.~Pricoupenko~\cite{pricoupenko1} that unlike the case of  the
strongly interacting bosonic  TG limit~\cite{KWW,paredes}, the
requirements for the observation of the fTG limit are quite
stringent,  and the effective range of scattering needs to be
taken into account. We provide an exact solution that accounts
for both scattering volume and effective range, and obtain
significant deviations
from the CS model~\cite{CSh} due to
``shape'' resonance in the p-wave scattering. We find several new
effects, such as the inversion of particle- and hole-like spectra
for low momenta, roton-like minima in the excitation spectrum,
and  we  calculate density profiles and ``breathing'' modes
 in the harmonic trap.

We use the asymptotic Bethe Ansatz (BA)~\cite{Sutherlandbook}
which is justified at sufficiently small densities, when only
two-particle collisions are important~\cite{gurarie}. The
underlying idea goes back  to the earlier days of high energy
physics and was known as S-matrix theory~\cite{Chew} in the
1950s. The BA method considers the scattering matrix between
asymptotic states as an alternative to a Hamiltonian/Lagrangean
description.
The scattering of ultracold atoms in 1D gases close  to resonance
can naturally be
described by the scattering 
phase shift, whereas the formulation of  a microscopic quantum
Hamiltonian is difficult. It can be  shown that the scattering
matrix close  to resonance corresponds to a highly singular,
although local, two-body interaction in the spirit of
Refs.~\cite{CSh, piter}. To avoid difficulties related to the
determination of the operators and states in this case,  we
use an approach based entirely on the scattering phase shift.


Let us start by briefly reviewing the 3D scattering properties in
a  p-wave channel. At low energies, the phase shift
$\delta_{p}(k)$ can be expanded as~\cite{pricoupenko1,LanLif}
 \bea\label{s-shift}
k^{3}\cot\delta_{p}(k)=-1/w_{1}-\alpha_{1}k^{2}+O(k^{4}),
 \eea
where $w_{1}$ is the scattering volume, $\alpha_{1}$ is the
effective range, and $k$ is the relative momentum.  For $\alpha_1
w_1<0,$
the scattering length obtained from $\delta_{p}(k)$ has a very
sharp  shape resonance at 
the wave vector~\cite{LanLif,pricoupenko2}
 \bea \label{krdef}
 k_r=1/\sqrt{-\alpha_1 w_1}.
 \eea
Such resonance is absent for s-wave scattering, but for  the
p-wave channel it exists  due to the presence of the effective
range parameter.  The higher order terms in Eq.~(\ref{s-shift})
do not significantly affect the shape resonance,  since they are
suppressed by powers of the small parameter $kR\ll 1$, where $R$
is the characteristic radius of the 3D potential.  For fermions,
the typical momenta of scattering particles are of the order of
the Fermi momentum $k_F=\pi n.$ Therefore, the condition $kR\ll
1$ necessary for neglecting three-particle
collisions~\cite{gurarie} implies the low-density limit $nR\ll 1$.
%
%
It is known~\cite{pricoupenko1,LanLif,pricoupenko2}  that
$\alpha_{1}\gtrsim 1/R
>0$ does not change significantly at the p-wave Feshbach
resonance, while $w_{1}$ can be tuned to very large absolute
values compared to its characteristic values of the order of
$|w_{1}|\sim R^{3}$ away from the resonance.

Under transverse harmonic confinement with frequency
$\omega_{\perp},$ only the lowest transverse mode is occupied if
the  momenta of scattering fermions satisfies
 \bea\label{ka-condition}
ka_\perp \ll 1,
 \eea  where $a_\perp=\sqrt{\hbar/(m\omega_{\perp})},$
 and $m$ is the atomic mass.
Under such conditions, the 1D scattering amplitude in an odd
channel is given by~\cite{pricoupenko1}
 \bea
 \label{f-odd} f_{p}^{odd}=\frac{-ik}{1/l_{p}+ik+k^{2}\xi_{p}},
 \eea
where~\cite{note}
 \bea
l_{p}=3a_{\perp}\left[\frac{a_{\perp}^{3}}{w_{1}}-3\sqrt{2}\zeta(-1/2)\right]^{-1},\quad
\xi_{p}=\frac{\alpha_{1}a_{\perp}^{2}}{3}>0,
 \eea
  and $3\sqrt{2}\zeta(-1/2)\approx -0.88.$ The notation adopted is that of Ref.~\cite{gurarie}.
  Using estimates~\cite{pricoupenko1,TicknorChevy} of $\alpha_{1}$ for $^{6}$Li and
$^{40}$K atoms at resonances with $B_{0}\approx 215$G and $
198.6$G, and  with transverse frequencies
$\omega_{\perp}=2\pi\times 200$kHz and $2\pi\times
30$kHz~\cite{Gunter}, we obtain $\xi_{p}/a_{\perp}\approx 50$ and
$\approx 13,$ respectively.
 Thus, under typical experimental conditions needed to achieve the
1D  regime $\xi_{p}\gg a_{\perp}$ and hence all three terms are
significant in the denominator of Eq.~(\ref{f-odd}).


The many-body fermionic wave function $\psi(z_{1},\ldots,z_{M})$
is anti-symmetric,
$\psi(\ldots,z_{i},\ldots,z_{j},\ldots)=-\psi(\ldots,z_{j},\ldots,z_{i},\ldots),$
and discontinuous when two coordinates coincide~\cite{CSh}.  We
define its symmetrized version by $
\psi_{+}(z_{1},\ldots,z_{M})=\prod_{i<j}\mbox{Sign}(z_{i}-z_{j})\psi(z_{1},\ldots,z_{M}),
$  which is  continuous. Then Eq.~(\ref{f-odd}) implies the
following boundary condition:
 \bea \lim_{z=z_{j}-z_{i}\rightarrow
0^{+}}\left(\frac{1}{l_{p}}+\partial_{z}-\xi_{p}\partial^{2}_{z}\right)\psi_{+}(z_{1},\ldots,z_{M})=0.\label{bc}
 \eea
  Solving the two-body problem as $\psi_+(z_{1},z_{2})\propto
e^{i\lambda|z_{1}-z_{2}|},$ we obtain two roots $
\lambda_{\pm}=\left(-i\pm\sqrt{-1-4\xi_{p}/l_{p}}\right)/(2\xi_{p}).
$
For $l_p > 0$,  Im$\lambda_{+} > 0$, which corresponds to a bound
state.  The  lowest energy state satisfying the boundary
condition (\ref{bc}) can then be constructed as  $
\psi(z_{1},\ldots,z_{M})\propto\prod_{i<j}\mbox{Sign}(z_{i}-z_{j})\prod_{i<j}\exp(i\lambda_{+}|z_{i}-z_{j}|).
 $  As in the attractive LL model, its energy does not have a
proper thermodynamic limit, we will not consider the case where
$l_p > 0$. For $l_{p} < 0$ we construct an exact wavefunction
$\psi_{+}(z_{1},\ldots,z_{M})$ as a combination of plane waves,
using  the BA method in a similar manner to the LL model.
In our case, such construction leads to
the following periodic boundary conditions on a circle of length
$L$
 \bea\label{eff-BAeq}
e^{i\lambda_{j}L}=\prod_{k=1}^{M}\frac{\xi_{p}(\lambda_{j}-\lambda_{k})^{2}-\frac{1}{|l_{p}|}+i(\lambda_{j}-\lambda_{k})}
{\xi_{p}(\lambda_{j}-\lambda_{k})^{2}-\frac{1}{|l_{p}|}-i(\lambda_{j}-\lambda_{k})},
 \eea
 and the total energy is given in terms of quasimomenta
$\lambda_{i}$ as $E=\hbar^{2}/(2m)\sum\lambda_{i}^{2}.$ We prove
that all  solutions of Eq.~(\ref{eff-BAeq}) are real by writing
the $k$-th term in the product as $\frac{(\lambda_j-\lambda_k
-\lambda_+)(\lambda_j-\lambda_k -\lambda_-)}{(\lambda_j-\lambda_k
-\lambda_+^*)(\lambda_j-\lambda_k -\lambda_-^*)}$. Since
Im$\lambda_{\pm}<0$  for $\xi_{p}>0$ and $l_{p}<0$, we then have
$\left|\frac{(\lambda
-\lambda_+)(\lambda-\lambda_-)}{(\lambda-\lambda_+^*)(\lambda-\lambda_-^*)}\right|\leq
1(\geq 1)$  for Im$\lambda\leq 0$$(\geq0)$. After that, the proof
simply follows the steps for the LL model described  on p.11 of
Ref.~\cite{KBI}.

To obtain a thermodynamic limit, we take a logarithm of Eq.
(\ref{eff-BAeq}), which is written as $L\lambda_{j}
+\sum_{k=1}^{M}\theta(\lambda_{j}-\lambda_{k})=2\pi n_{j},$  where
$n_j$ are integer quantum numbers for odd $M.$ The phase shift
$\theta(\lambda)$ is a monotonic antisymmetric function defined by
 \bea
\theta(\lambda)=2\mbox{Arg}(i\lambda-\xi_{p}\lambda^{2}+1/|l_{p}|),
 \eea
  and  belongs to the interval $(-2\pi,2\pi),$ unlike the LL phase  shift, which belongs to the interval $(-\pi,\pi).$ We then
directly follow Ref.~\cite{YY} and show that real solutions of
the BA equations exist for any choice of quantum numbers $n_{j}.$
Their values for the ground state can be fixed by comparison with
the LL model~\cite{LL, KBI,YY}, and are  given by $n_{j} =
j-(M+1)/2.$ Introducing a positive function
 \bea
K(\lambda,\mu)=\theta^\prime(\lambda,\mu)=\frac{2|l_{p}|\left[1+|l_{p}|\xi_{p}(\lambda-\mu)^{2}\right]}{\left[1-|l_{p}|\xi_{p}(\lambda-\mu)^{2}\right]^{2}+l_{p}^{2}(\lambda-\mu)^{2}}
\nonumber,
 \eea
we pass to the thermodynamic limit, and write an equation for the
ground state quasimomenta distribution in the  usual way
$2\pi\rho(\nu)-\int_{-q}^{q}K(\nu,\mu)\rho(\mu)d\mu=1$, where
$\pm q$ is the highest~(lowest) filled quasimomentum and  the
normalization is given by $n = M/L =\int_{-q}^{q}\rho(\nu)d\nu$.
Apart from new definitions of $\theta(\lambda)$ and
$K(\lambda,\mu),$ the  structure of the theory is similar to the
LL model, and we can study the  ground state energy, excitation
spectra and finite temperature properties using standard
methods~\cite{KBI}.

\begin{figure}
\includegraphics[width=8.5cm]{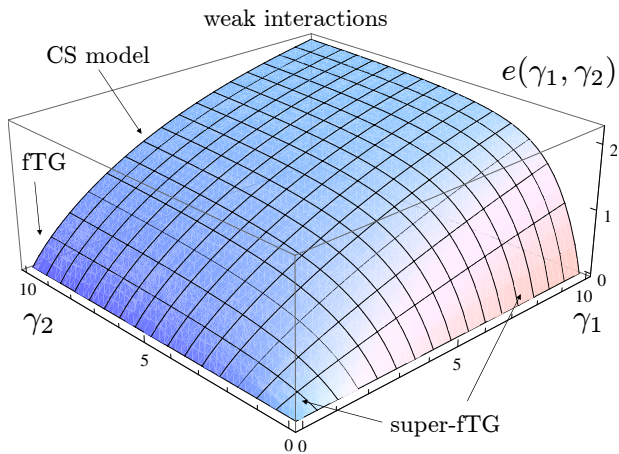}
\vspace{-0.5cm}
 \caption{\label{Fig1} (Color online) Energy
functional $e(\gamma_1,\gamma_2)$ as a function of dimensionless
parameters $\gamma_1$ and $\gamma_2,$ see
Eqs.~(\ref{gammadef})-(\ref{edef}). Arrows indicate the regimes
of weakly interacting fermions, Cheon-Shigehara (CS)
model~\cite{CSh}, fermionic Tonks-Girardeau (fTG) gas~\cite{GW},
and super-fTG regime (see Eq.~(\ref{sftg})) existing due to the
finite effective range $\xi_p.$} \vspace{-0.5cm}
\end{figure}

We choose two dimensionless parameters that determine the ground
state properties in the stable region
 \bea \gamma_{1}=-\frac{1}{l_{p}n}>0,\quad
\gamma_{2}=\frac{1}{\xi_{p}n}>0 \label{gammadef}. \eea In
Fig.~\ref{Fig1} we show the dimensionless ground state energy
functional $e(\gamma_{1},\gamma_{2})$ obtained by numerically
solving the equations for the ground state. The dimensionless
form is given by the expression
 \bea
 E/L=e(\gamma_{1},\gamma_{2})(\hbar n)^{2}/(2m), \label{edef}
 \eea  and reduces to the LL functional $e(\gamma_1)$ for
$\gamma_2\gg 1,$ since the CS model~\cite{CSh} obtained in this
limit is dual to the LL model. The function
$e(\gamma_{1},\gamma_{2})$ equals $0$ if $\gamma_{1}=0$ or
$\gamma_{2}=0.$
Expansion by methods of Ref.~\cite{Wadati} at $\gamma_1,\gamma_2\gg 1$ yields
 $e(\gamma_{1},\gamma_{2})\approx
e(\gamma_1)-32\pi^4/(15\gamma_1^2\gamma_2).$

\begin{figure}
\includegraphics[width=8.5cm]{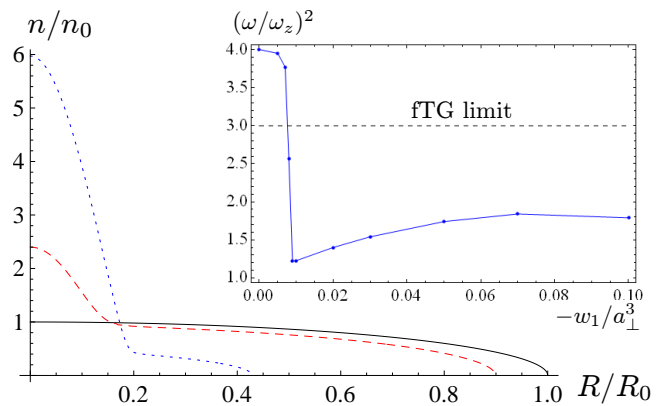}
 \vspace{-0.5cm}
\caption{\label{Fig2} (Color online) Density profiles under
harmonic confinement, measured in units of Thomas-Fermi radius
$R_0$ and central density $n_0$ for the same cloud in the absence
of interactions. Different curves correspond to
$\{\gamma_1(0),\gamma_2(0)\}$ in the center of the cloud $ \{7.5,
0.023\}$(blue, dotted), $\{97, 0.058\}$(red, dashed),  and  free
fermions (black, thick). For a cloud of $N=100$ $^6$Li atoms with
$\omega_\perp=2\pi\times 200\;\mbox{kHz}$ and
$\omega_z=2\pi\times 200\;\mbox{Hz},$ we get $R_0=41\mu{\rm m}$
and these  curves correspond to $-w_1/a_\perp^3=0.05, 0.01, 0,$
respectively. The inset shows the ratio of squares  of the
``breathing'' and dipole mode
frequencies $(\omega/\omega_z)^2$ 
for the same trap parameters. If a significant part of the cloud
is in the  super-fTG regime (see text), this ratio drops below
the critical value $3$ obtained in the fTG limit. A  critical
value $-w_1/a_{\perp}^3\approx 0.008$ corresponds to  a detuning
 from the resonance $\Delta B\approx 50$mG.}
 \vspace{-0.5cm}
\end{figure}

We use  the function $e(\gamma_{1},\gamma_{2})$ obtained
numerically to evaluate density profiles in a harmonic trap
within the local density approximation~\cite{LDA}.  We also use
the function $e(\gamma_{1},\gamma_{2})$ to find the ``breathing''
mode frequency $\omega$  by solving the hydrodynamic
equations~\cite{mixture1,Chiara}.
 The results are shown in
Fig.~\ref{Fig2} and  depend on two dimensionless parameters,
$\gamma_1(0)$ and $\gamma_2(0),$ in the center of the cloud.
The presence of the effective range strongly affects the shape of
the profile compared to the CS model if $\gamma_2(0)$ is small and
$\gamma_1(0)$ is not too large.  This effect can be understood by
using the expansion of $e(\gamma_1,\gamma_2)$ for
$\gamma_1,\gamma_2 \ll 1.$ The leading term in the Taylor
expansion gives $e(\gamma_1,\gamma_2)\propto
\gamma_1\gamma_2\propto 1/n^2$ and hence via Eq.~(\ref{edef}) the
energy per particle for this term does not depend on density,
i.e. the  gas has a divergent compressibility. Higher order terms
in the  expansion 
lead to a finite but
large compressibility, which decreases with increasing $\gamma_1$
and a constant ratio $\gamma_2/\gamma_1.$ Thus, the density
profile exhibits a strong peak near the  center where $\gamma_1$
is smallest.

 Since fermions are more strongly correlated for
$\gamma_2\ll 1$ and not too large $\gamma_1$ compared to the fTG
regime, we suggest calling such a regime a super-fTG gas,
analogously to the super-TG gas of bosons~\cite{STG}.
%
%
%
%
It is realized if
 \bea
 \mbox{Max}\; (\gamma_2, \frac{\gamma_1 \gamma_2}{4\pi^2}) \lesssim
 1.\label{sftg}
 \eea
  For $|w_1|\ll a^3_\perp,$ the second condition corresponds to
 $
 k_r \lesssim 2\pi n = 2k_F.
 $
 Thus, the super-fTG regime is realized if the largest relative
momentum of non-interacting fermions approaches the shape
resonance wave vector $k_r.$ In a similar manner to the super-TG
gas of bosons, the  super-fTG regime can be experimentally
identified by measuring the ratio  of the squares  of the
``breathing'' and dipole mode frequencies. In the  CS model such
a ratio is always larger  than $3$, similar to the LL
model~\cite{Chiara}, while the inset of Fig.~\ref{Fig2} shows the
regime where it is smaller than $3.$ A sharp decrease in this
ratio for the super-fTG regime can be easily understood from the
``sum rule'' approach of Ref.~\cite{Chiara}, since the cloud
density is much more centered in the super-fTG regime than in the
fTG regime. We can analytically estimate the value of
$-w_1/a_\perp^3$ at which the center of the cloud enters the
super-fTG regime, and the drop in $(\omega/\omega_z)^2$ occurs.
For that we use Eq.~(\ref{sftg}) with the free fermion density in
the center and obtain $
  -\frac{w_1}{a_\perp^3}\Big|_{\rm super-fTG}=\frac{\omega_{\perp}/\omega_z}{24 N
  \xi_p/a_{\perp}},
 $
which gives $0.008$ for the parameters seen in Fig.~\ref{Fig2}.

\begin{figure}
\includegraphics[width=8.5cm]{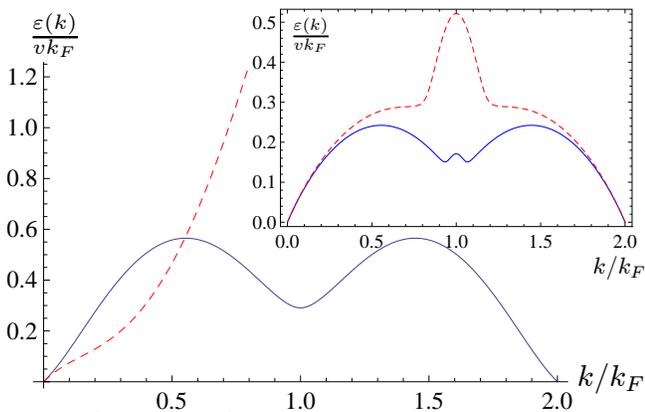}

 \vspace{-0.5cm}

\caption{\label{Fig3} (Color online) Excitation spectrum for
$\gamma_1=0.5, \gamma_2=0.005,$ where momentum is measured in
units of $k_F= \pi n,$ and energy is normalized to give a unit
velocity $v=\varepsilon'(k)|_{k=0}.$ Due to interactions, the
particle-like mode (red, dashed) has energy lower than the
hole-like mode  (blue, solid) for sufficiently small momenta.
Velocities of two modes at $k=0$ coincide. The inset shows
energies of hole-like modes for $\gamma_1=10.8, \gamma_2=0.022$
(blue, solid) and $\gamma_1=27.3, \gamma_2=0.055$  (red, dashed),
normalized to their respective velocities.} \vspace{-0.5cm}
\end{figure}

 The excitation spectrum $\varepsilon(k)$ in a uniform cloud is also significantly modified in the super-fTG regime compared to
predictions of the CS model. In Fig.~\ref{Fig3} we illustrate
several qualitative features, which appear due to the finite
effective range of interactions. Firstly, the system has a regime
where the energy of the particle-like excitation is smaller than
the energy of the hole-like excitation. Since the energy of the
particle-like excitation should  approach $k^2/(2\, m)$ at high
momenta, there should also be  an energy crossing. This crossing
will manifest itself as a kink in the $k-$dependence of the
lowest energy of the density wave excitations. Secondly, the
spectrum of  hole-like excitations can have a ``roton'' minimum
(or even an additional maximum, see inset in Fig.~\ref{Fig3}) at
$k\approx k_F.$ This minimum can be understood as a tendency of
the system towards pairing  when the parameters $\gamma_1,
\gamma_2$ approach the boundary of the stable region. Indeed, the
energy of the hole-like excitation vanishes for $k=2k_F=2 \pi n.$
Since particle density is twice the density of pairs, the
vicinity of the paired region manifests itself as a soft mode at
$k\approx2 \pi n/2=k_F.$  Dynamic response functions of the
system will have power-law divergences at  the particle- and
hole-like modes, which can be calculated using the methods of
Ref.~\cite{LLexp}.  Note however, that the existence of the roton
minimum and the inversion of the particle- and hole-like spectra
lead to modifications of the phenomenology of
Ref.~\cite{universal}.


We thank E.~Demler and C.~Bolech for discussions at  the early
stages of this work, and R.~Hulet for useful comments. This
research was supported by NSF DMR Grant No. 0906498 and by the
Swiss NSF.

\begin{thebibliography}{99}


\bibitem{1}
S.~Inouye {\it et al.}, Nature (London) {\bf 392}, 151 (1998);
Ph.~Courteille {\it et al.}, Phys. Rev. Lett. {\bf 81}, 69
(1998); J.L.~Roberts {\it et al.}, {\it ibid.} {\bf 81}, 5109
(1998).

\bibitem{KWW}
T.~Kinoshita, T.~Wenger, and D.S.~Weiss, Science {\bf 305}, 1125
(2004).

\bibitem{paredes}
B.~Paredes {\it et al.}, Nature (London) {\bf 429}, 277 (2004).

\bibitem{LL}
E.H.~Lieb and W.~Liniger, Phys. Rev. {\bf 130}, 1605 (1963);
E.H.~Lieb, {\it ibid.} {\bf 130}, 1616 (1963).

\bibitem{FRZ}
J.N.~Fuchs, A.~Recati, and W.~Zwerger, Phys. Rev. Lett. 93, 090408
(2004); I.V.~Tokatly, {\it ibid.} {\bf 93}, 090405 (2004);
 G.~Orso, {\it ibid.} {\bf 98}, 070402 (2007); H.~Hu,
X.-J.~Liu, and P.D.~Drummond, {\it ibid.} {\bf 98}, 070403 (2007);
L.~Guan {\it et al.}, {\it ibid.} {\bf 102}, 160402 (2009).


\bibitem{mixture1}
A.~Imambekov and E.~Demler, Phys. Rev. A {\bf 73}, 021602(R)
(2006); Ann. Phys. {\bf 321}, 2390 (2006).

\bibitem{mixture2}
M.T.~Batchelor {\it et al.}, Phys. Rev. A {\bf 72}, 061603(R)
(2005); H.~Frahm and G.~Palacios, {\it ibid.} {\bf 72}, 061604(R)
(2005).

\bibitem{spinor_b}J.~Cao, Y.~Jiang, and Y.~Wang, EPL {\bf 79}, 30005 (2007);
F.~Deuretzbacher {\it et al.}, Phys. Rev. Lett. {\bf 100}, 160405
(2008).
\bibitem{spinor_f}Y.~Jiang, J.~Cao, and Y.~Wang, EPL {\bf 87}, 10006 (2009).



\bibitem{pricoupenko1}
L. Pricoupenko, Phys. Rev. Lett. {\bf 100}, 170404 (2008).

\bibitem{p-wave1}
C.A.~Regal {\it et al.,}
Phys. Rev. Lett. {\bf 90}, 053201 (2003); J.P.~Gaebler {\it et
al.,} 
{\it ibid.} {\bf 98}, 200403 (2007).

\bibitem{p-wave2}
J.~Zhang {\it et al.}, Phys. Rev. A {\bf 70}, 030702(R) (2004);
C.H.~Schunck {\it et al.}, {\it ibid.} {\bf 71}, 045601 (2005);
J.~Fuchs {\it et al.}, {\it ibid.} {\bf 77}, 053616 (2008);
Y.~Inada {\it et al.}, Phys. Rev. Lett. {\bf 101}, 100401 (2008).

\bibitem{Gunter}K.~G\"{u}nter {\it et al.}, Phys. Rev. Lett. {\bf 95}, 230401 (2005).

\bibitem{3D}V.~Gurarie, L.~Radzihovsky, and A.V.~Andreev, Phys. Rev. Lett. {\bf 94}, 230403
(2005); C.-H.~Cheng and S.-K.~Yip, {\it ibid.} {\bf 95}, 070404
(2005); J.~Levinsen, N.R.~Cooper, and V.~Gurarie, {\it ibid.}
{\bf 99}, 210402 (2007).

\bibitem{GB}
B.E.~Granger and D.~Blume, Phys. Rev. Lett. {\bf 92}, 133202
(2004).

\bibitem{CSh}
T.~Cheon and T.~Shigehara, Phys. Rev. Lett. {\bf 82}, 2536 (1999);
Phys. Lett. A {\bf 243}, 111 (1998).

\bibitem{GW}
M.D.~Girardeau and E.M.~Wright, Phys. Rev. Lett. {\bf 95}, 010406
(2005); S.A.~Bender, K.D.~Erker, and B.E.~Granger, {\it ibid.}
{\bf 95}, 230404 (2005); M.D.~Girardeau and A.~Minguzzi, {\it
ibid.} {\bf 96}, 080404 (2006).

\bibitem{Sutherlandbook}B. Sutherland, {\it Beautiful models} (World Scientific, Singapore, 2004).

\bibitem{gurarie}
V. Gurarie, Phys. Rev. A {\bf 73}, 033612 (2006).


\bibitem{Chew}
S.C. Frautschi, {\it Regge poles and S-matrix theory} (Benjamin,
New York, 1963).






\bibitem{piter}
V.S.~Buslaev  and N.A.~Kaliteevsky, Teor. Math. Phys. {\bf 70},
 187 (1987).


\bibitem{LanLif}
L.D.~Landau and E.M.~Lifshitz, {\it Quantum Mechanics}
(Butterworth-Heinemann, Oxford, 1981), $\S133,$ p.552.

\bibitem{pricoupenko2}
L.~Pricoupenko, Phys. Rev. A {\bf 73}, 012701 (2006); Phys. Rev.
Lett. {\bf 96}, 050401 (2006).

\bibitem{note}Our $a_{\perp}$ is $\sqrt{2}$ times smaller than that of
Ref.~\cite{pricoupenko1}.


\bibitem{TicknorChevy}C. Ticknor {\it et al.}, Phys. Rev. A {\bf 69}, 042712 (2004); F. Chevy {\it et al.,} {\it ibid.} {\bf 71}, 062710 (2005).



\bibitem{KBI}
V.E.~Korepin, N.M.~Bogoliubov and A.G.~Izergin, {\it Quantum
Inverse Scattering Method and Correlation Functions} (Cambridge
University Press, Cambridge, England, 1993).


\bibitem{YY}
C.N. Yang and C.P. Yang, J. Math. Phys. {\bf 10}, 1115 (1969).

\bibitem{Wadati}T.~Iida and M.~Wadati, J. Phys. Soc. Jpn. {\bf 74},
1724 (2005).

\bibitem{LDA} V.~Dunjko, V.~Lorent, and  M.~Olshanii, Phys. Rev. Lett. {\bf 86}, 5413
(2001).
\bibitem{Chiara}C.~Menotti and S.~Stringari, Phys. Rev. A {\bf 66}, 043610
(2002).

\bibitem{STG}G.E.~Astrakharchik {\it et al.,} 
Phys. Rev. Lett. {\bf 95}, 190407 (2005); E.~Haller {\it et al.},
Science {\bf 325}, 1224 (2009).

\bibitem{LLexp} A.~Imambekov and L.I.~Glazman, Phys. Rev. Lett. {\bf 100}, 206805 (2008).

\bibitem{universal}A.~Imambekov and L.I.~Glazman, Science {\bf 323}, 228 (2009);  Phys.
  Rev. Lett. {\bf 102}, 126405 (2009).


\end{thebibliography}
\end{document}